\documentclass[prl,aps,twocolumn,floatfix,superscriptaddress]{revtex4-1}

\pdfoutput=1
\usepackage{amsmath,amssymb}
\usepackage{bm,bbm}
\usepackage{graphicx}
\usepackage{mathrsfs}
\usepackage{slashed}
\usepackage{booktabs}
\usepackage{tabu}
\usepackage[dvipsnames]{xcolor}
\usepackage[normalem]{ulem}
\usepackage{tikz}
\usepackage{soul}
\usetikzlibrary{shapes,arrows}
\usetikzlibrary{positioning}
\usepackage[colorlinks=true,linkcolor=blue]{hyperref}
\usepackage{xcolor}
\usepackage{ulem}

\begin{document}

\preprint{JLAB-THY-18-2678}

\title{First Monte Carlo global QCD analysis of pion parton distributions}

\author{P. C. Barry}
\affiliation{North Carolina State University,
	     Raleigh, North Carolina 27607, USA}
\author{N. Sato}
\affiliation{University of Connecticut,
	     Storrs, Connecticut 06269, USA}
\author{W. Melnitchouk}
\affiliation{Jefferson Lab,
	     Newport News, Virginia 23606, USA \\
        \vspace*{0.2cm}
        {\bf Jefferson Lab Angular Momentum (JAM) Collaboration
        \vspace*{0.2cm} }}
\author{Chueng-Ryong Ji}
\affiliation{North Carolina State University,
	     Raleigh, North Carolina 27607, USA}

\begin{abstract}
We perform the first global QCD analysis of parton distribution
functions (PDFs) in the pion, combining $\pi A$ Drell-Yan data with
leading neutron electroproduction from HERA within a Monte Carlo
approach based on nested sampling.
Inclusion of the HERA data allows the pion PDFs to be determined down
to much lower values of $x$, with relatively weak model dependence
from uncertainties in the chiral splitting function.
The combined analysis reveals that gluons carry a significantly
higher pion momentum fraction, $\sim 30\%$, than that inferred from
Drell-Yan data alone, with sea quarks carrying a somewhat smaller
fraction, $\sim 15\%$, at the input scale.
Within the same effective theory framework, the chiral splitting
function and pion PDFs can be used to describe the $\bar d-\bar u$
asymmetry in the proton.
\end{abstract}

\date{\today}
\maketitle

%
As the lightest QCD bound state, the pion has historically played
a central role in the study of the strong nuclear interactions.
On one hand, it has been the critical ingredient for understanding
the consequences of dynamical chiral symmetry breaking in QCD, and
how this dictates the nature of hadronic interactions at low energies.
On the other hand, its quark and gluon (or parton) substructure has
been revealed through high energy scattering experiments, such as
Drell-Yan (DY) lepton-pair creation in inclusive pion--nucleon
scattering \cite{DY}.
In some cases, both aspects are on display, as in the role of
the pion cloud of the proton in generating a flavor asymmetry in
its light antiquark sea, $\bar d \neq \bar u$~\cite{Thomas83}.

As the simplest $q \bar q$ state, the structure of the pion is
relatively more straightforward to compute theoretically than
baryons, but the absence of fixed pion targets has made it
difficult to determine the pion's structure experimentally.
Most information on the partonic structure of pions has come from
pion--nucleus scattering with prompt photon or dilepton production
at CERN~\cite{NA3, NA10} and Fermilab \cite{E615}.
These data have been used in several QCD analyses
\cite{Owens84, ABFKW89, SMRS, GRV, GRS, Wijesooriya05, Aicher10}
to fit the momentum dependence of the pion's parton distribution
functions (PDFs) for parton fractions $x_\pi \gtrsim 0.1$ of the
pion's light-cone momentum.

While the DY data constrain reasonably well the pion's valence PDFs,
the sea quark and gluon PDFs at small $x_\pi$ values have remained
essentially unknown.
More recently, leading neutron (LN) electroproduction from HERA
\cite{ZEUS, H1}, which at forward angles is expected to be dominated
by pion exchange, has been used to study the pion structure function
down to very low values of $x_\pi \sim 10^{-3}$.
The interpretability of the LN data in terms of pion PDFs is limited,
however, by the model dependence inherent in this process, in which
the cross section is given as a product of a proton to neutron
$+$ pion ``chiral splitting function'' and the structure function
of the (nearly on-shell) exchanged pion.
Consequently the LN data have never been used in QCD analyses,
although recently the first steps toward their inclusion were taken
by McKenney {\it et~al.} \cite{LN15}, who studied the impact of
the model dependence on the extracted pion structure function by
constraining the $p \to n \pi^+$ splitting function empirically.

In the present work, we combine the strategy of global QCD analysis
with an empirical approach to using the DY and LN data in the same fit
to determine the pion PDFs in both the high-$x_\pi$ and low-$x_\pi$
regions.
We use for the first time a Monte Carlo (MC) approach, based on the
nested sampling algorithm \cite{Skilling, MPL08, Shaw}, to perform
the global analysis at next-to-leading order (NLO) in the strong
coupling.
In contrast to the previous single-fit analyses based on maximum
likelihood methods \cite{Owens84, ABFKW89, SMRS, GRV, GRS,
Wijesooriya05, Aicher10}, the MC approach allows a systematic
exploration of the parameter space by computing the likelihood
function directly, providing a rigorous determination of PDF
uncertainties.

An important feature of our analysis is the ability of the MC fit to
quantify the uncertainty on the dependence of the extracted pion PDFs
on the chiral splitting function model.
To test the robustness of the chiral framework, we also perform a
simultaneous fit to the pion DY $+$ LN data together with the E866
$pd/pp$ DY data \cite{E866}, from which the $\bar d/\bar u$ ratio
was extracted, using the same MC methodology.
Such an analysis provides the most comprehensive study of pion PDFs
and their impact on different observables.

%
In the pion-induced DY process \cite{DY}, partons from the pion
and target nucleus $A$ annihilate to produce a dimuon pair in the
final state, $\pi A \to \mu^+ \mu^- X$, with cross section
\begin{multline}
\frac{d^2\sigma}{dQ^2 dY}
= \frac{4\pi \alpha^2}{9 Q^2 S}
  \sum_{i, j}
  \int_{x_\pi}^1 \frac{d\hat{x}_\pi}{\hat{x}_\pi}
  \int_{x_A}^1   \frac{d\hat{x}_A}{\hat{x}_A}		\\
  \times C_{ij}(\hat{x}_\pi,\hat{x}_A,x_\pi,x_A,Q/\mu)\,
	 f_i^\pi(\hat{x}_\pi,\mu)\,
	 f_j^A(\hat{x}_A,\mu),
\label{eq.DYxsec}
\end{multline}
where $f_i^\pi$ ($f_j^A$) is the PDF for parton flavor $i$ in the pion
(flavor $j$ in the nucleus) as a function of the parton momentum
fraction $\hat{x}_\pi$ ($\hat{x}_A$), and $C_{ij}$ is the hard
scattering kernel \cite{BNX, ADMP}, with $\mu$ the renormalization
scale.
The cross section is differential in the dilepton invariant mass
squared $Q^2$ and rapidity $Y$, in terms of which one defines
	$x_{\pi,A} = \sqrt{\tau}\, e^{\pm Y}$,
where $\tau = Q^2/S$ and $S$ is the $\pi$-target invariant mass squared.
At lowest order the parton momentum fractions are given by
	$\hat{x}_{\pi, A} = x_{\pi, A}$.
Typically, the experimental DY cross sections are analyzed in terms
of the Feynman variable $x_F = x_\pi - x_N$, where $x_N = A x_A$ is
the nuclear Bjorken variable scaled per nucleon \cite{DS03}.
%
%
The available pion DY data from the CERN NA10~\cite{NA10} and Fermilab
E615 \cite{E615} experiments were taken on tungsten nuclei, and for the
nuclear PDFs in our analysis we use the parametrization from Eskola
{\it et al.} \cite{EPPS16}.

%
For the LN production process, $e p \to e n X$, the pion PDFs enter
indirectly, under the assumption that the charge exchange cross section
at low values of $t$ and large neutron longitudinal momentum fractions
$x_L$ is dominated by single pion exchange.  The differential LN
cross section $d^3 \sigma^{\rm LN}/dx dQ^2 dx_L$ is parametrized by
the LN structure function, $F_2^{\rm LN(3)}(x,Q^2,x_L)$.
According to the chiral effective theory of QCD, matching twist-two
partonic and corresponding hadronic operators leads then to a
factorized representation of $F_2^{\rm LN(3)}$ \cite{ChenPRL, 
Moiseeva13, XWangPRD},
\begin{equation}
F_2^{\rm LN(3)}(x,Q^2,x_L)
= 2 f_{\pi N}(\bar{x}_L)\, F_2^\pi(x_\pi, Q^2).
\label{eq.F2LN}
\end{equation}
Here $f_{\pi N}(\bar{x}_L)$ is the chiral splitting function for
fraction $\bar{x}_L \equiv 1 - x_L = x/x_\pi$ of the proton's
light-cone momentum carried by the pion, and $F_2^\pi(x_\pi,Q^2)$
is the pion structure function, evaluated at NLO.
%
%
The splitting function is evaluated from chiral effective theory
\cite{Z1, Burkardt13, Salamu15}, and for $\bar{x}_L > 0$ is given by
\begin{eqnarray}
f_{\pi N}(\bar{x}_L)
&=& \frac{g_A^2 M^2}{(4 \pi f_\pi)^2}
    \int\!dk_\perp^2
    \frac{\bar{x}_L \left[ k_\perp^2 + \bar{x}_L^2 M^2 \right]}
	 {x_L^2\, D_{\pi N}^2}\,
    |{\cal F}|^2,
\label{eq.splfnc}
\end{eqnarray}
where 
$D_{\pi N} \equiv t-m_\pi^2
 = -[k_\perp^2 + \bar{x}_L^2 M^2 + x_L m_\pi^2]/x_L$,
with $M$ and $m_\pi$ the nucleon and pion masses,
$g_A$ the axial charge, and $f_\pi$ the pion decay constant.
The form of the splitting function in Eq.~(\ref{eq.splfnc}) is
constrained by chiral symmetry in QCD \cite{Z1, Ji09, Comment}, and its
infrared or leading nonanalytic behavior is model independent
\cite{TMS00, Detmold01, Chen01, Arndt02}.
The ultraviolet behavior, however, is dependent on the regularization
procedure, represented in Eq.~(\ref{eq.splfnc}) by the function
${\cal F}$.
In the literature various forms have been advocated, including cutoff
regularization, Pauli-Villars, and phenomenological $\pi N$ form
factors, and following Ref.~\cite{LN15} we consider several forms,
\begin{equation}
{\cal F} = \left\{
\begin{array}{lcll}
&\hspace*{-0.35cm}\text{(i)}&
 \hspace*{-0.05cm}\exp\left( (M^2-s)/\Lambda^2 \right)
	& s\text{-dep.\! exponential}		\\
&\hspace*{-0.35cm}\text{(ii)}&
 \hspace*{-0.05cm}\exp\left( D_{\pi N}/\Lambda^2 \right)
	& t\text{-dep.\! exponential}		\\
&\hspace*{-0.35cm}\text{(iii)}&
 \hspace*{-0.05cm}(\Lambda^2-m_\pi^2)/(\Lambda^2-t)
	& t\text{-dep.\! monopole}		\\
&\hspace*{-0.35cm}\text{(iv)}&
 \hspace*{-0.05cm}\bar{x}_L^{-\alpha_\pi(t)}
	  \exp\left( D_{\pi N}/\Lambda^2 \right)
	& \text{Regge}				\\
&\hspace*{-0.35cm}\text{(v)}&
 \hspace*{-0.05cm}\left[ 1 - D_{\pi N}^2/(\Lambda^2-t)^2 \right]^{1/2}
	& \text{Pauli-Villars}			\nonumber
\label{eq.models}
\end{array}
\right.
\end{equation}
where
$s = (k_\perp^2+M^2) / x_L + (k_\perp^2+m_\pi^2) / \bar{x}_L$,
$\alpha_\pi(t) = \alpha'_\pi t$,
with $\alpha'_\pi \approx 1$~GeV$^{-2}$,
and $\Lambda$ is a cutoff parameter.
%
%
We also considered a model based on a large-$k_\perp$ cutoff
\cite{Salamu15}, and the Bishari model~\cite{ZEUS, Bishari}
(which is analogous to the Regge form but with $\Lambda \to \infty$).
While these also give reasonable descriptions of the (low-$t$)
LN data, they become problematic for observables that are more
sensitive to large $t$, such as the $\bar d-\bar u$ asymmetry
from E866 \cite{E866}.

In addition to the LN structure function data from H1~\cite{H1},
the ZEUS collaboration measured the ratio \cite{ZEUS}
\begin{eqnarray}
r(x,Q^2,x_L)
&=& \frac{d^3 \sigma^{\rm LN}/dxdQ^2dx_L}{d^2\sigma^{\rm inc}/dxdQ^2}
    \Delta x_L
\label{eq.r}
\end{eqnarray}
of LN to inclusive cross sections, where the latter is expressed in
terms of the proton structure function, $F_2^p$, and $\Delta x_L$
is the bin size in $x_L$.
Consistent with expectations from earlier theoretical calculations
\cite{DAlensio00, Kopeliovich12}, at large $x_L \sim 1$ pion exchange
is the dominant contribution~\cite{LN15}.  Other processes, such as
absorption and the exchange of other mesons, play an increasingly
important role at smaller $x_L$.
Instead of choosing a specific minimum value of $x_L$ above which
one pion exchange is assumed, we fit the minimum value of $x_L$
for which the data can be described within this framework.


For the data analysis we use a Bayesian Monte Carlo method based
on the nested sampling algorithm \cite{Skilling, MPL08, Shaw},
which allows a faithful Monte Carlo representation of the probability
distribution 
  ${\cal P}({\bm a}|{\rm data})
  = {\cal L}({\rm data}|{\bm a}) \pi({\bm a}) / Z$,
where ${\bm a}$ is an $n$-dimensional array of the pion PDFs shape 
parameters. 
Here $\pi(\bm{a})$ is the Bayesian prior distribution for $\bm{a}$,
which allows the parameter sampling to be restricted to physical
regions,
  ${\cal L}({\rm data}|{\bm a})
  = {\rm exp}[-\frac{1}{2} \chi^2({\bm a})]$
is the likelihood function, and
  $Z = \int d^{n}a\, {\cal L}({\rm data}|{\bm a})\, \pi({\bm a})$
is the Bayesian evidence parameter, which normalizes the probability
distribution.  
We use a $\chi^2$ function in the likelihood that takes into account
correlated systematic shifts, as well as overall normalizations of
the data sets~\cite{MSTW}.
For physical observables $\cal O$, such as the pion PDFs and functions
thereof, from the MC samples $\{ \bm{a}_k \}$ one then obtains
expectation values
  ${\rm E}[{\cal O}]
  = \sum_k w_k\, {\cal O}(\bm{a}_k)$
and variances
  ${\rm V}[{\cal O}]
  = \sum_k w_k \left({\cal O}(\bm{a}_k) - {\rm E}[{\cal O}]\right)^2$
where
$\{ w_k \}$ are the MC weights.
Similar MC technology based on Bayesian statistics has also been
applied recently to study nucleon PDFs \cite{Sato16, Ethier17} and
fragmentation functions \cite{SatoFF16}, as well as the transverse
momentum dependent transversity distribution \cite{Lin18}.

For the pion valence PDFs we assume charge symmetry,
$q_v^\pi \equiv u_v^{\pi^+} = u^{\pi^+} - {\bar u}^{\pi^+}
	 = \bar{d}_v^{\pi^+}
	 = \bar{u}_v^{\pi^-}
	 = d_v^{\pi^-}$,
and invoke SU(3) symmetry for the pion sea,
$q_s^\pi \equiv {\bar u}^{\pi^+}
	 = d^{\pi^+} = s^{\pi^+} = {\bar s}^{\pi^+}$.
The valence, sea quark, and gluon PDFs are parameterized at the input
scale of the charm quark mass $Q_0^2 = m_c^2 = (1.3~\rm GeV)^2$
by the form
\begin{equation}
f(x_\pi, Q_0^2; \bm{a})
= \frac{N}{B(2+\alpha,\beta)}\, x_\pi^\alpha (1-x_\pi)^\beta,
\label{eq.PDF}
\end{equation}
where $\bm{a} = \{ N, \alpha, \beta \}$ are the fitting parameters
and $B$ is the Euler beta function.
The valence PDFs are normalized such that
    $\int_0^1 dx_\pi\, q_v^\pi = 1$,
and the momentum sum rule gives the constraint
    $\int_0^1 dx_\pi x_\pi\, ( 2 q_v^\pi + 6 q_s^\pi + g^\pi ) = 1$.

\begin{figure}[t]
\hspace*{-0.2cm}\includegraphics[width=0.5\textwidth]{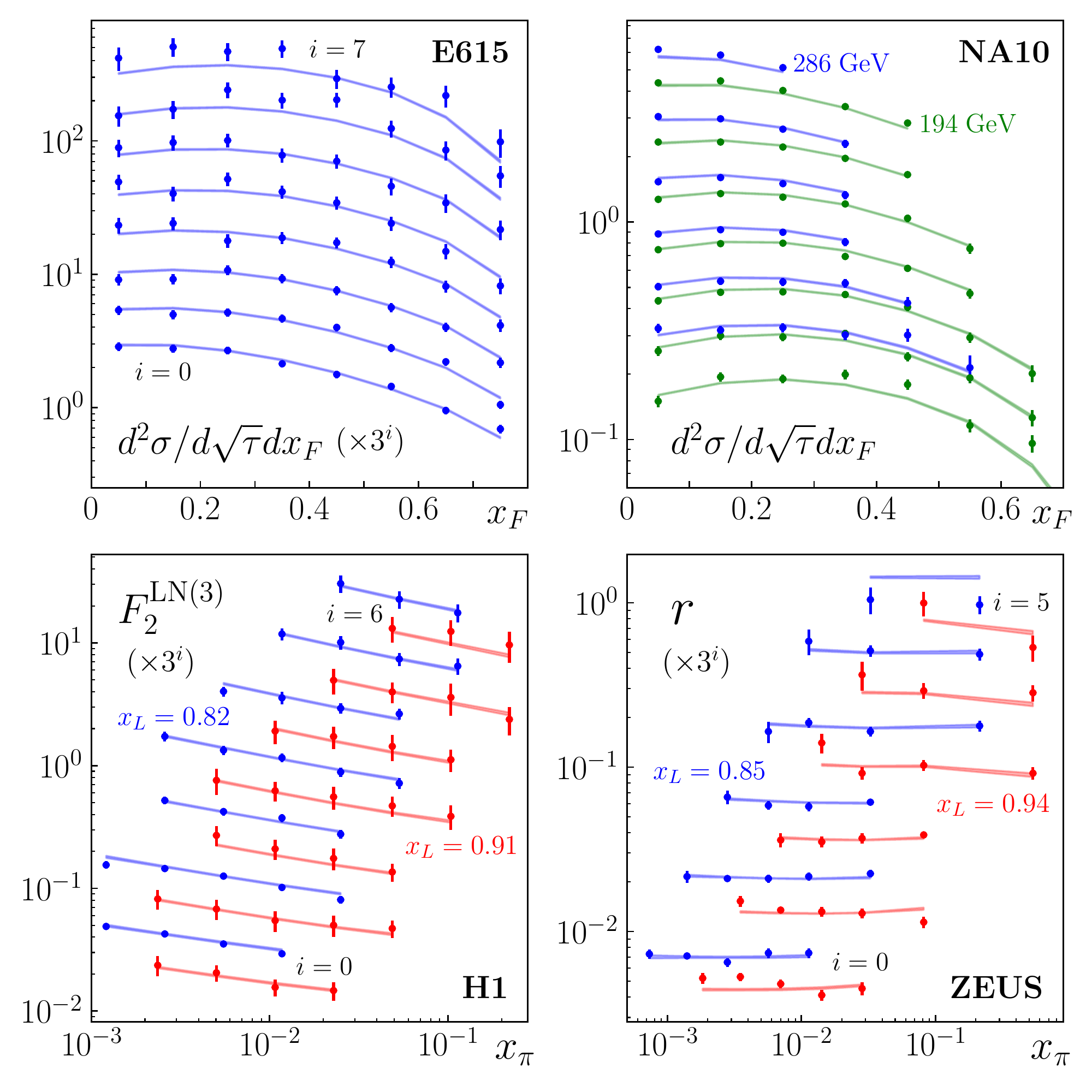}
\vspace*{-0.8cm}
\caption{Cross sections computed with our fitted pion PDFs
	compared with DY $d^2\sigma/d\sqrt\tau dx_F$ data
	from E615~\cite{E615} {\it (top left)} and
	NA10~\cite{NA10} {\it (top right)} [in units of nb],
	and with the LN structure function $F_2^{\rm LN(3)}$ from
	H1~\cite{H1} {\it (bottom left)} and LN to inclusive ratio $r$
	from ZEUS~\cite{ZEUS} {\it (bottom right)}.
	For display purposes, the E615, H1 and ZEUS data
	are scaled by a factor $3^i$ for clarity.
	The NA10 data are for $\pi^-$ beam energies of
	194~GeV (green) and 286~GeV (blue).}
\label{f.datonthy} 
\end{figure}

The fits to the DY and LN data sets are shown in Fig.~\ref{f.datonthy},
where for clarity the E615 and HERA points are scaled by $3^i$.
To avoid the $J/\Psi$ and $\Upsilon$ resonances, the DY data were
restricted to the mass region
    $4.16 < Q < 8.34~{\rm GeV}$,
covering the range
    $0.05 \leq x_F \leq 0.9$.
Generally very good agreement is found for the entire set of 250
data points. 
%
%
For the best fit, corresponding to model (i) for the LN cross section
with a cutoff $\Lambda = 1.31(4)$~GeV, the combined $\chi^2/N_{\rm dat}$
is 0.98 (244.8/250).
%
Increasing the number of parameters in Eq.~(\ref{eq.PDF}) did not
improve the overall $\chi^2$.
%
The overall normalizations for the DY data are found to be
	0.816, 0.758 and 0.985
for the	NA10 (194~GeV), NA10 (286~GeV) and E615 data sets, and
	1.17 and 0.964
for the H1 and ZEUS LN data, respectively.

For the LN data good fits were obtained for the cut $x_L > 0.8$;
including smaller-$x_L$ data deteriorated the fit due to larger
non-pionic contributions away from the forward limit~\cite{LN15,
DAlensio00, Kopeliovich12}.
One could extend the region over which to fit the data by including
also non-pionic contributions, such as from vector or axial vector
mesons, though this would be at the expense of introducing more
parameters into the analysis and diluting the connection with QCD.
Fitting the DY data alone yields only marginally smaller $\chi^2$
values, with $\chi^2/N_{\rm dat} = 0.97$ (55.5/70 for NA10 and
82.6/72 for E615).
For the combined DY and LN data sets, the total $\chi^2$ per datum
for other models are also close to 1.0, with
$\chi^2 = \{ 267.7,\, 266.0,\, 262.8,\, 273.8 \}$ for models (ii)--(v),
corresponding to cutoff parameters
$\Lambda = \{ 0.58(2),\, 0.52(2),\, 0.78(5),\, 0.25(1) \}$~GeV,
respectively.

\begin{figure}[t]
\hspace*{-0.2cm}\includegraphics[width=0.5\textwidth]{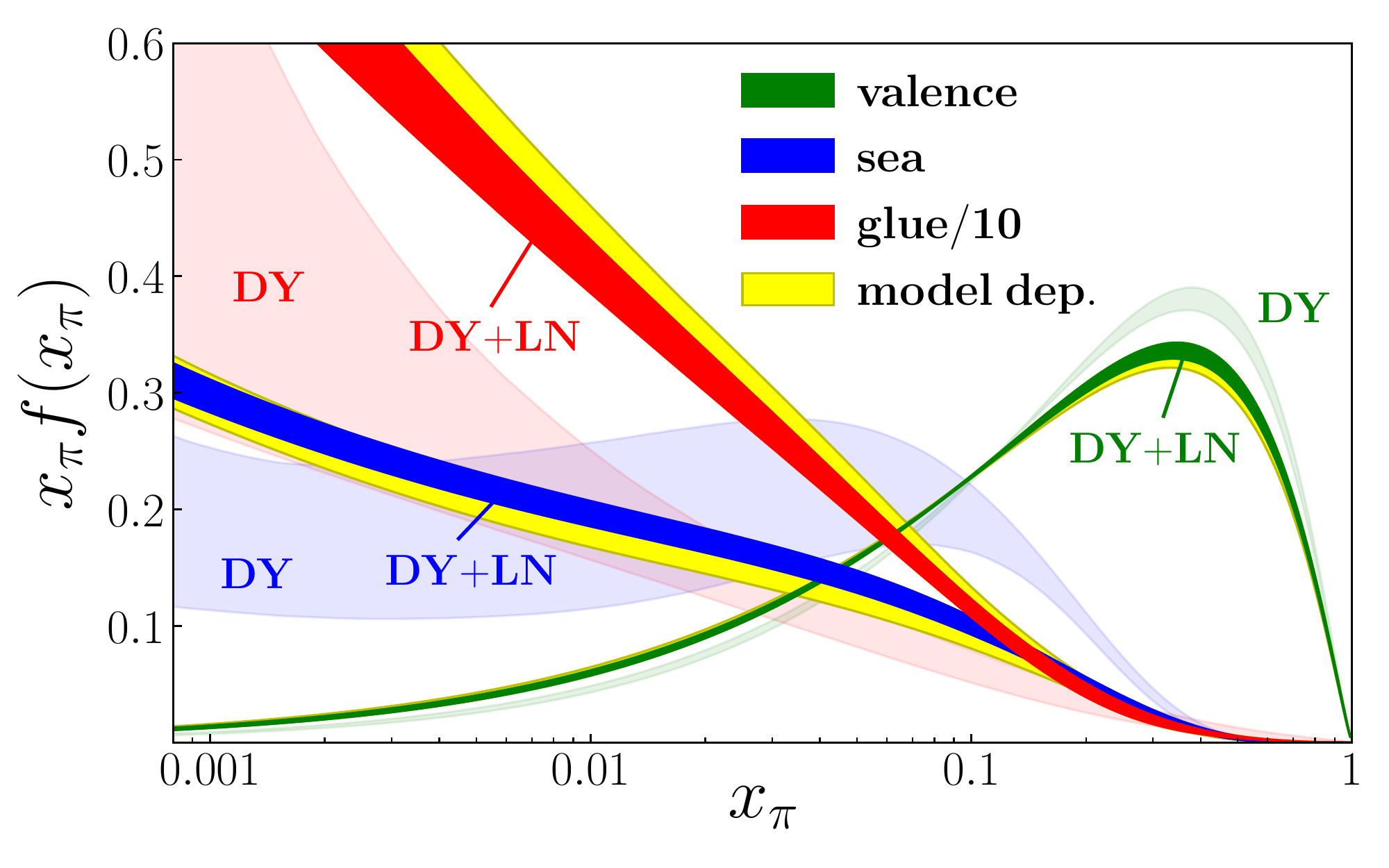}
\vspace*{-0.8cm}
\caption{Pion valence (green), sea quark (blue) and gluon (red,
	scaled by 1/10) PDFs versus $x_\pi$ at $Q^2=10$~GeV$^2$,
	for the full DY $+$ LN (dark bands) and DY only
	(light bands) fits.  
	The bands represent 1$\sigma$ uncertainties,
	as defined in the standard Monte Carlo determination of the
	uncertainties~\cite{SatoFF16} from the experimental errors.
	The model dependence of the fit is represented by the
	outer yellow bands.}
\label{f.pdf} 
\end{figure}

The resulting pion PDFs are shown in Fig.~\ref{f.pdf} for the
valence, sea quark, and gluon distributions at $Q^2 = 10$~GeV$^2$.
Compared with the DY-only fits, which constrain mainly the valence
quark PDF and for $x_\pi \lesssim 0.1$ are essentially an extrapolation,
the simultaneous DY$+$LN analysis yields significantly reduced
uncertainties on the pion sea and gluon distributions at low $x_\pi$.
While the addition of the LN data gives a small, $\approx 10\%$
reduction of the valence PDF at intermediate $x_\pi$, the impact on
the sea is more dramatic, with the gluon PDF increasing twofold at
$x_\pi \sim 0.001-0.1$ compared with the DY-only result, but with
half of the uncertainty.
The sea quark PDF $q_s^\pi$ is reduced at $x_\pi \gtrsim 0.1$,
but is slightly larger at $x_\pi \lesssim 0.1$ for the full result.
Importantly, the model dependence of the combined DY+LN fit
(represented in Fig.~\ref{f.pdf} by the yellow bands) reveals
a relatively small uncertainty, especially compared with the
scale of the effect induced by the addition of the LN results.

For the valence PDF in the large-$x_\pi$ region, our analysis
finds a behavior $\sim (1-x)$ at the input scale, which is
harder than expectations based on pQCD \cite{pQCD} which
prefer a $(1-x)^2$ fall-off.
Expectations from low energy models vary in their estimates of the
$x \to 1$ behavior \cite{Toki93, Arriola95, Szczepaniak94, DSE,
dual_pi, AdS}, and generally the scale at which these are applicable
is not clear.
Furthermore, the present analysis does not include threshold
resummation effects, which are known to be important at large $x_\pi$
\cite{Aicher10, Westmark17}, and this will be examined in a separate
analysis \cite{Barry18}.

\begin{figure}[t]
\hspace*{-0.2cm}\includegraphics[width=0.5\textwidth]{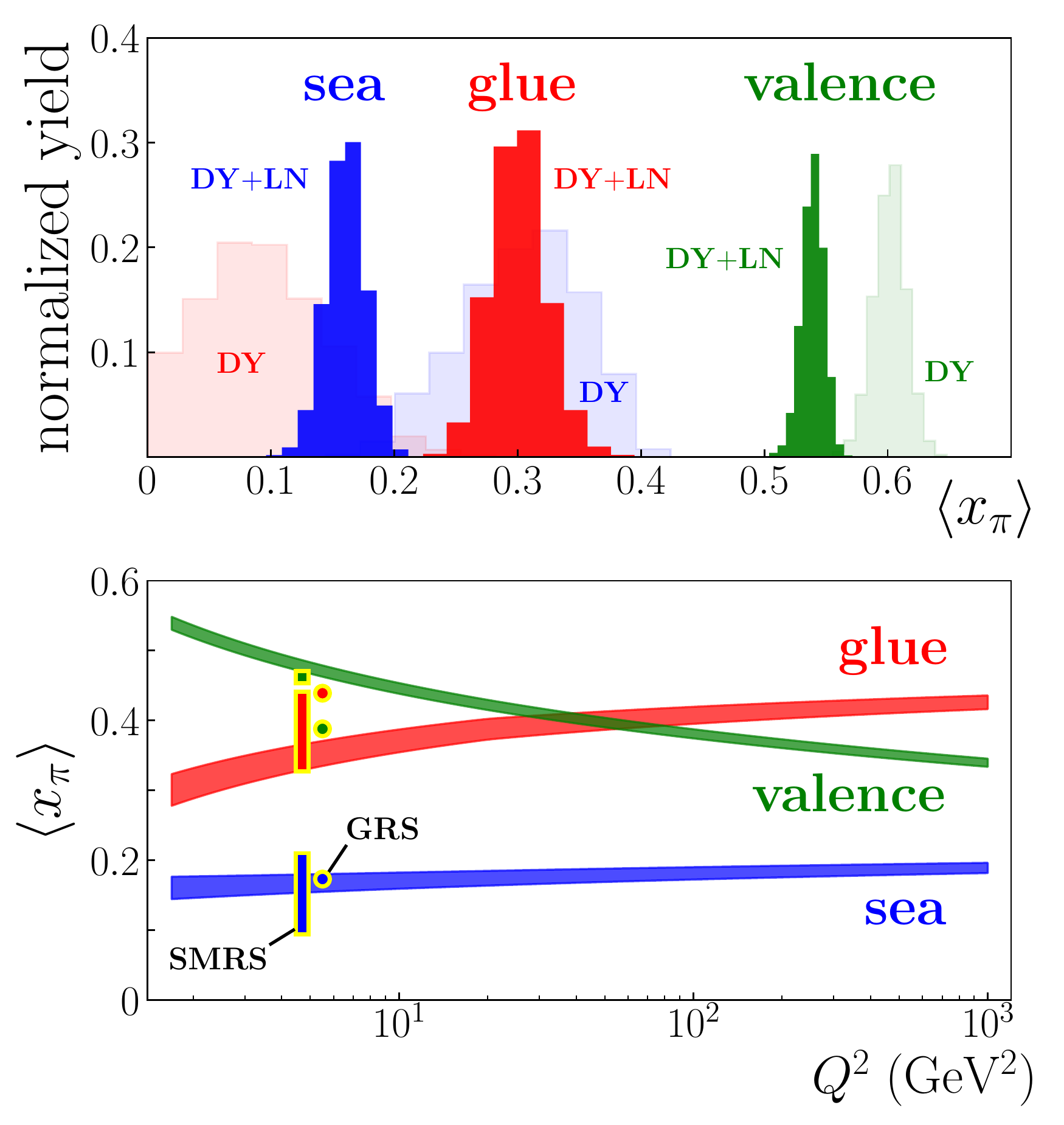}
\vspace*{-0.8cm}
\caption{{\it (Top)}\ Normalized yield for the average momentum
	fractions $\langle x_\pi \rangle$ of the pion carried by
	valence quarks (green),	sea quarks (blue) and gluons (red)
	at $Q^2 = m_c^2$, for the full MC fit of DY+LN data
	(dark shaded) and for DY-only (light shaded).
	{\it (Bottom)}\ Scale dependence of the momentum fractions
	for the full DY+LN fit, compared with results from the
	SMRS (rectangular bands)~\cite{SMRS} and GRS~\cite{GRS}
	(circles) parametrizations at $Q^2 = 5$~GeV$^2$
	(offset for clarity).}
\label{f.msr} 
\end{figure}

The inclusion of the LN data into the global analysis allows a more
precise breakdown of the pion momentum into fractions carried by
valence quarks, sea quarks, and gluons, shown in Fig.~\ref{f.msr}.
The total valence, sea and gluon momentum fractions for the full
analysis at the input scale $Q^2=m_c^2$ are found to be
	$\{ \langle x_\pi \rangle_v,
	    \langle x_\pi \rangle_s,
	    \langle x_\pi \rangle_g \}$
	$= \{ 54(1)\%,\, 16(2)\%,\, 30(2)\% \}$
for the best fit with model~(i).
Compared with the DY-only fit, where the respective momentum fractions
are $\{ 60(1)\%,\, 30(5)\%,\, 10(5)\% \}$, the fraction carried by gluons
is about 3 times larger, but with less than half of the uncertainty.
Since the valence fraction remains relatively unchanged, the momentum
sum rule forces the sea quarks to carry about 1/2 of the momentum
fraction compared with the DY-only fit.

This turns out to be similar to the result from the older SMRS analysis
\cite{SMRS} of DY plus prompt-photon data, which considered several
scenarios in which the momentum fraction carried by sea quarks at
$Q^2 = 5$~GeV$^2$ varied from 10\% to 20\%.
With the valence momentum fraction in \cite{SMRS} constrained to be
46\%, the momentum fraction carried by gluons varied from 43\% to 34\%
over this range.
At the same scale the analogous fractions from our analysis are
	$\{ 48(1)\%,\, 17(1)\%,\, 35(2)\% \}$,
which is closer to the SMRS scenario with maximum sea and minimum glue.
In contrast, the GRS analysis~\cite{GRS}, which also fits DY + prompt
photon data, assuming a constituent quark model to constrain the sea
quark and gluon PDFs, gives a gluon momentum fraction (44\%) that is
actually larger than the valence fraction (39\%) at this scale.
%

%

\begin{figure}[t]
\hspace*{-0.2cm}\includegraphics[width=0.51\textwidth]{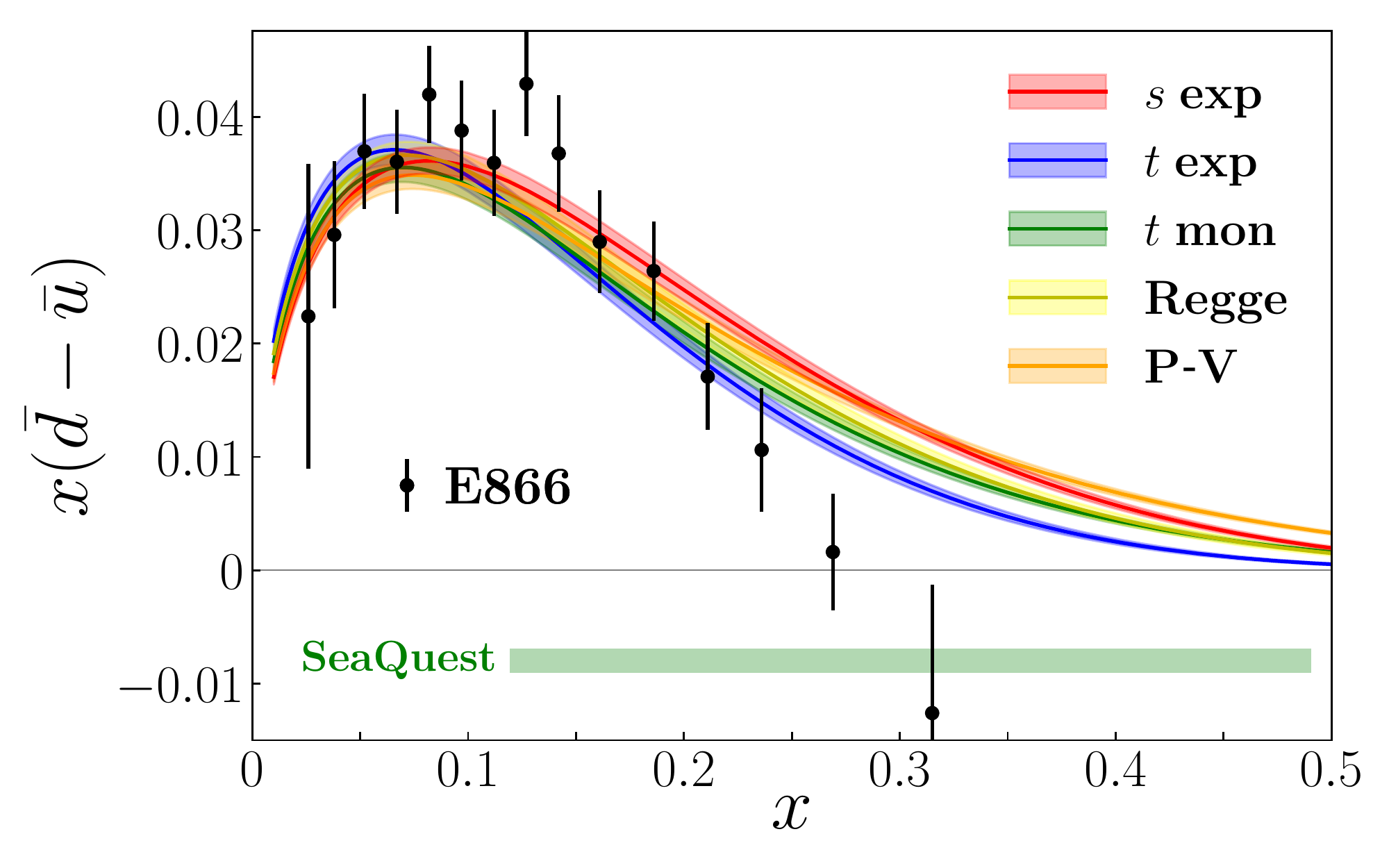}
\vspace*{-0.8cm}
\caption{$\bar{d}-\bar{u}$ asymmetry in the proton for various
	chiral splitting function models, with parameters from
	a combined fit to the $\pi A$ DY~\cite{E615, NA10},
	LN production~\cite{ZEUS, H1} and E866~\cite{E866}
	data (black circles) at $Q^2=54$~GeV$^2$.
	The kinematic coverage of the SeaQuest experiment
	\cite{SeaQuest} is indicated by the horizontal band.}
\label{f.E866} 
\end{figure}

As an application of our results and a test of the validity of the
chiral framework for the LN data, we consider the contribution from the
$p \to n\, \pi^+$ dissociation to the $\bar{d} - \bar{u}$ asymmetry
in the proton sea.
This asymmetry, predicted from pion loop effects in the nucleon
\cite{Thomas83}, was conclusively established by the $pp$ and $pd$
Drell-Yan data from the E866 experiment at Fermilab \cite{E866}.
While the effect of these data on PDFs is most rigorously quantified
through global QCD fits, an approximate NLO analysis of the E866
data extracted the $\bar{d} - \bar{u}$ difference, shown in
Fig.~\ref{f.E866}, assuming knowledge of the valence quark
distributions in the proton.

The antiquark asymmetry can be represented within chiral effective
theory as a convolution (represented by the symbol ``$\otimes$'')
of the $p \to n\, \pi^+$ and $p \to \Delta^{++}\, \pi^-$ splitting
functions and the valence pion PDF \cite{ChenPRL, Moiseeva13,
XWangPRD, SMST},
  \mbox{$\bar{d}-\bar{u}$}
  $= (f_{\pi^+ n} - \frac23 f_{\pi^- \Delta^{++}}) \otimes q_v^\pi$.
The $\pi \Delta$ splitting function, which does not contribute to
the LN cross section but enhances the $\bar u$ PDF over $\bar d$
for inclusive processes, is given in Ref.~\cite{Salamu15}.
Refitting the DY and LN data together with the 15 additional E866
extracted data points, and attributing the entire asymmetry to pion
loops, gives a best fit for model~(i) with only a marginally larger
$\chi^2/N_{\rm dat} = 1.03$ (272.4/265).
The combined fit has additional sensitivity to the valence quark
pion PDF, however, the resulting PDFs are relatively stable, and
the fitted model~(i) cutoff parameter, $\Lambda = 1.35(2)$~GeV,
is consistent with that from the fit without E866 data.

The resulting $\bar d - \bar u$ asymmetry is shown in Fig.~\ref{f.E866},
for all models~(i)--(v) for the regulator ${\cal F}$.
Without compromising the description of the DY or LN data, a reasonably
good fit to the E866 data points can be achieved in all cases for
$x \lesssim 0.2$, beyond which all the fits overestimate the data.
It is known that the apparent change of sign in $\bar d - \bar u$ at
high $x$ is difficult to accommodate theoretically~\cite{Speth98},
and the new DY SeaQuest experiment~\cite{SeaQuest} will allow a more
precise determination of the asymmetry up to $x \approx 0.5$.

%
%
%


In the future, pion PDFs will be further constrained by new $\pi A$
DY data from COMPASS~\cite{COMPASS, FrancoCOMPASS}, as well as from
the Tagged DIS (TDIS)~\cite{TDIS} experiment at Jefferson Lab,
which will study pion structure through the charge exchange mechanism
in leading proton production form a quasi-free neutron in the deuteron,
$e d \to e p p X$.
%
The new data may shed light on the lack of overlap between the DY-only
and DY+LN fits in the valence region, as may future analyses with
non-Gaussian likelihoods to further investigate possible tensions among
data sets.
%
Theoretically, effects from gluon resummation \cite{Aicher10,
Westmark17} and higher twists \cite{pQCD} will be explored
\cite{Barry18} systematically in order to unravel the behavior
of pion PDFs at very high $x_\pi \sim 1$.
Beyond this, an ultimate future goal will be a simultaneous fit of
pion, proton and nuclear PDFs within the same MC global QCD analysis.

%
This work was supported by the U.S. Department of Energy (DOE)
Contract No.~DE-AC05-06OR23177, under which Jefferson Science
Associates, LLC operates Jefferson Lab, by NSFC under Grant
No.~11475186, CRC 110 by DFG and NSFC,
and by DOE Contract No. DE-FG02-03ER41260.



\begin{thebibliography}{99} 

\bibitem{DY}
S.~D.~Drell and T.-M.~Yan,
Phys. Rev. Lett. {\bf 25}, 316 (1970)
[Erratum ibid. {\bf 25}, 902 (1970)].

\bibitem{Thomas83}
A.~W.~Thomas,
Phys. Lett. B {\bf 126}, 97 (1983).

\bibitem{NA3}
J.~Badier {\it et al.},
Z. Phys. C {\bf 18}, 281 (1983).

\bibitem{NA10}
B.~Betev {\it et al.},
Z. Phys. C {\bf 28}, 9 (1985).

\bibitem{E615}
J.~S.~Conway {\it et al.},
Phys. Rev. D {\bf 39}, 1 (1989).

\bibitem{Owens84}
J.~F.~Owens,
Phys. Rev. D {\bf 30}, 943 (1984).

\bibitem{ABFKW89}
P.~Aurenche, R.~Baier, M.~Fontannaz, M.~N.~Kienzle-Focacci and M.~Werlen,
Phys. Lett. B {\bf 233}, 517 (1989).

\bibitem{SMRS}
P.~J.~Sutton, A.~D.~Martin, W.~J.~Stirling, R.~G.~Roberts,
Phys. Rev. D {\bf 45}, 2349 (1992).

\bibitem{GRV}
M.~Gl\"{u}ck, E.~Reya and A.~Vogt,
Z. Phys. C {\bf 53}, 651 (1992).

\bibitem{GRS}
M.~Gl\"{u}ck, E.~Reya and I.~Schienbein,
Eur. Phys. J. C {\bf 10}, 313 (1999).

\bibitem{Wijesooriya05}
K.~Wijesooriya, P.~E.~Reimer and R.~J.~Holt,
Phys. Rev. C. {\bf 72}, 065203 (2005).

\bibitem{Aicher10}
M.~Aicher, A.~Sch\"{a}fer and W.~Vogelsang,
Phys. Rev. Lett. {\bf 105}, 252003 (2010).

\bibitem{ZEUS}
S.~Chekanov {\it et al.}, 
Nucl. Phys. {\bf B637}, 3 (2002).

\bibitem{H1}
F.~D.~Aaron {\it et al.}, 
Eur. Phys. J. C {\bf 67}, 381 (2010).

\bibitem{LN15}
J.~R.~McKenney, N.~Sato, W.~Melnitchouk and C.-R.~Ji,
Phys. Rev. D {\bf 93}, 054011 (2016).

\bibitem{Skilling}
J.~Skilling,
Bayesian Anal. {\bf 1}, 4 (2006).

\bibitem{MPL08}
P.~Mukherjee, D.~Parkinson and A.~Liddle,
Astrophys. J. {\bf 638}, L51 (2006).

\bibitem{Shaw}
R.~Shaw, M.~Bridges and M.~P.~Hobson,
Mon. Not. Roy. Astron. Soc. {\bf 378}, 1365 (2007).

\bibitem{E866}
R.~S.~Towell {\it et al.},
Phys. Rev. D {\bf 64}, 052002 (2001).

\bibitem{BNX}
T.~Becher, M.~Neubert and G.~Xu,
JHEP {\bf 0807}, 030 (2008).

\bibitem{ADMP}
C.~Anastasiou, L.~Dixon, K.~Melnikov and F.~Petriello,
Phys. Rev. Lett. {\bf 91}, 182002 (2003).

\bibitem{DS03}
D.~de~Florian and R.~Sassot,
Phys. Rev. D {\bf 69}, 074028 (2004).

\bibitem{EPPS16}
K.~Eskola, P.~Paakkinen, H.~Paukkunen and C.~Salgado,
Eur. Phys. J. C {\bf 77}, 163 (2017).

\bibitem{ChenPRL}
J.-W.~Chen and X.~Ji,
Phys. Rev. Lett. {\bf 87}, 152002 (2001);
{\bf 88}, 249901(E) (2002).

\bibitem{Moiseeva13}
A.~M.~Moiseeva and A.~A.~Vladimirov,
Eur. Phys. J. A {\bf 49}, 23 (2013). 

\bibitem{XWangPRD}
X.~Wang, C.-R.~Ji, W.~Melnitchouk, Y.~Salamu, A.~W.~Thomas and P.~Wang,
Phys. Rev. D {\bf 94}, 094035 (2016).

\bibitem{Z1}
C.-R.~Ji, W.~Melnitchouk and A.~W.~Thomas,
Phys. Rev. D {\bf 88}, 076005 (2013).

\bibitem{Burkardt13}
M.~Burkardt, K.~S.~Hendricks, C.~R.~Ji, W.~Melnitchouk and A.~W.~Thomas,
Phys. Rev. D {\bf 87}, 056009 (2013).

\bibitem{Salamu15}
Y.~Salamu, C.-R.~Ji, W.~Melnitchouk and P.~Wang,
Phys. Rev. Lett. {\bf 114}, 122001 (2015).

\bibitem{Ji09}
C.-R.~Ji, W.~Melnitchouk and A.~W.~Thomas,
Phys. Rev. D {\bf 80}, 054018 (2009).

\bibitem{Comment}
C.-R.~Ji, W.~Melnitchouk and A.~W.~Thomas,
Phys. Rev. Lett. {\bf 110}, 179101 (2013).

\bibitem{TMS00}
A.~W.~Thomas, W.~Melnitchouk and F.~M.~Steffens,
Phys. Rev. Lett. {\bf 85}, 2892 (2000).

\bibitem{Detmold01}
W.~Detmold, W.~Melnitchouk, J.~W.~Negele, D.~B.~Renner and A.~W.~Thomas,
Phys. Rev. Lett. {\bf 87}, 172001 (2001).

\bibitem{Chen01}
J.-W.~Chen and X.~Ji,
Phys. Lett. B {\bf 523}, 107 (2001).

\bibitem{Arndt02}
D.~Arndt and M.~J.~Savage,
Nucl. Phys. {\bf A697}, 429 (2002).

\bibitem{Bishari}
M.~Bishari,
Phys. Lett. {\bf 38B}, 510 (1972). 

\bibitem{DAlensio00}
U.~D'Alesio and H.~J.~Pirner,
Eur. Phys. J. A {\bf 7}, 109 (2000).

\bibitem{Kopeliovich12}
B.~Z.~Kopeliovich, I.~K.~Potashnikova, B.~Povh and I.~Schmidt,
Phys. Rev. D {\bf 85}, 114025 (2012).

\bibitem{MSTW}
L.~A.~Harland-Lang, A.~D.~Martin, P.~Motylinski and R.~S.~Thorne,
Eur. Phys. J. {\bf C} 75, (2015) 204.

\bibitem{Sato16}
N.~Sato, W.~Melnitchouk, S.~E.~Kuhn, J.~J.~Ethier and A.~Accardi,
Phys. Rev. D {\bf 93}, 074005 (2016).

\bibitem{Ethier17}
J.~J.~Ethier, N.~Sato and W.~Melnitchouk,
Phys. Rev. Lett. {\bf 119}, 132001 (2017).

\bibitem{SatoFF16}
N.~Sato, J.~J.~Ethier, W.~Melnitchouk, M.~Hirai, S.~Kumano and A.~Accardi,
Phys. Rev. D {\bf 94}, 114004 (2016).

\bibitem{Lin18}
H.-W.~Lin, W.~Melnitchouk, A.~Prokudin, N.~Sato and H.~Shows,
Phys. Rev. Lett. {\bf 120}, 152502 (2018).

\bibitem{pQCD}
G.~R.~Farrar and D.~R.~Jackson,
Phys. Rev. Lett. {\bf 43}, 246 (1979);
%
E.~L.~Berger and S.~J.~Brodsky,
Phys. Rev. Lett. {\bf 42}, 940 (1979);
%
S.~J.~Brodsky and F.~Yuan,
Phys. Rev. D {\bf 74}, 094018 (2006);
%
F.~Yuan,
Phys. Rev. D {\bf 69}, 051501 (2004).

\bibitem{Toki93}
T.~Shigetani, K.~Suzuki and H.~Toki,
Phys. Lett. B {\bf 308}, 383 (1993).

\bibitem{Szczepaniak94}
A.~Szczepaniak, C.-R.~Ji and S.~R.~Cotanch,
Phys. Rev. D {\bf 49}, 3466 (1994).

\bibitem{Arriola95}
R.~M.~Davidson and E.~Ruiz Arriola,
Phys. Lett. {\bf B}, 348 (1995).

\bibitem{DSE}
M.~B.~Hecht, C.~D.~Roberts and S.~M.~Schmidt,
Phys. Rev. C {\bf 63}, 025213 (2001).

\bibitem{dual_pi}
W.~Melnitchouk,
Eur. Phys. J. A {\bf 17}, 223 (2003);
Phys. Rev. D {\bf 67}, 077502 (2003).

\bibitem{AdS}
G.~de~T\'{e}ramond, T.~Liu, R.~S.~Sufian, H.~G.~Dosch, S.~J.~Brodsky and A.~Deur,
Phys. Rev. Lett. {\bf 120}, 182001 (2018).

\bibitem{Westmark17}
D.~Westmark and J.~F.~Owens,
Phys. Rev. D {\bf 95}, 056024 (2017).

\bibitem{Barry18}
P.~C.~Barry {\it et al.},
in preparation.

\bibitem{SMST}
A.~W.~Schreiber, P.~J.~Mulders, A.~I.~Signal and A.~W.~Thomas,
Phys. Rev. D {\bf 45}, 3069 (1992).

\bibitem{Speth98}
W.~Melnitchouk, J.~Speth and A.~W.~Thomas,
Phys. Rev. D {\bf 59}, 014033 (1998).

\bibitem{SeaQuest}
Fermilab E906 Experiment (SeaQuest),
{\it Drell-Yan Measurements of Nucleon and Nuclear Structure
with the Fermilab Main Injector},
{\tt http://www.phy.anl.gov/mep/ SeaQuest/index.html}.

\bibitem{COMPASS}
The Drell-Yan project at COMPASS,\\
{\tt http://wwwcompass.cern.ch/compass/future\char`_physics/ drellyan/}.

\bibitem{FrancoCOMPASS}
C.~Franco,
on behalf of the COMPASS collaboration,
XIV Hadron Physics, Florianopolis, March 2018,
{\it Polarised Drell-Yan results from COMPASS}.

\bibitem{TDIS}
J.~Annand, D.~Dutta, C.~E.~Keppel, P.~King and B.~Wojtsekhowski
(spokespersons), Jefferson Lab experiment E12-15-006.

\end{thebibliography}
\end{document}